\documentclass[%
 reprint,
 amsmath,amssymb,
 aps,
prb,
]{revtex4-2}

\usepackage{graphicx}
\usepackage{booktabs}
\usepackage{dcolumn}
\usepackage{bm}

\usepackage{tikz}
\usetikzlibrary{
    arrows.meta,
    positioning,
    calc,
    decorations.pathmorphing,
    decorations.markings,
    patterns
}
\begin{document}


\title{Engineering Weak Universality with Quantum Dots}

\author{Warre Missiaen}
 \email{w.missiaen@tudelft.nl}

\author{Michael Wimmer}%
\affiliation{%
 QuTech and Kavli Institute of Nanoscience, Delft University of Technology,
P.O. Box 4056, 2600 GA Delft, The Netherlands\\ \vspace{-0.4cm}
}%
\author{Natalia Chepiga}%
\affiliation{Rudolf Peierls Centre for Theoretical Physics, University of Oxford, Clarendon Laboratory, Oxford OX1 3PU, United Kingdom\\
}%

\date{\today}

\begin{abstract}
Quantum critical theories with continuously varying critical exponents remain challenging to access experimentally. Here, we propose quantum-dot architectures for realizing the quantum Ashkin–Teller and XYZ/eight-vertex models using resonator-mediated and direct Coulomb interactions, respectively. We focus on the Ashkin–Teller and eight-vertex critical lines, which are connected by a non-local duality relating local order parameters to topological string operators. The resulting platforms provide microscopic control over the parameters explicitly controlling critical exponents, with the Coulomb-based implementation enabling stronger interaction regimes. We demonstrate  that continuously varying critical behavior can already be resolved in short chains. For experimentally realistic parameters, the accessible interaction range is
expected to produce changes in the critical exponents of order $10\%$ for the
Ashkin--Teller model and substantially larger changes for the eight-vertex model.
In both cases, the predicted variations lie well above the estimated measurement
uncertainties.
\end{abstract}

\maketitle

\section{\label{sec:level1}Introduction}
Among one-dimensional quantum critical systems, theories with central charge $c=1$ constitute a particularly rich and intriguing class, encompassing the remarkable phenomenon of weak universality, in which critical exponents vary continuously along a critical line. The most prominent microscopic realizations of weak universality are provided by the quantum Ashkin--Teller (AT) \cite{Ashkin1943,DiFrancesco1997,giamarchi2004quantum} and XYZ models realizing the eight-vertex critical line on a special manifold of its exchange couplings \cite{PhysRevLett26834,DiFrancesco1997}.  In both cases, critical exponents vary continuously along $c=1$ critical lines \cite{Baxter1982,PhysRevLett26832,Chepiga_2023}. These critical lines are related by partial duality - an exact non-local transformation \cite{PhysRevA81032334,FENDLEY1989549}, bridging  two complementary descriptions of this continuously varying criticality.
\\

The quantum AT model consists of two coupled transverse-field Ising chains, with the interaction strength parametrizing a continuous line of $c=1$ fixed points \cite{aoun2023phasediagramashkintellermodel}. Along this line, the scaling dimension of the order parameter remains fixed at $d_{AT}=\beta_{\rm AT}/\nu_{\rm AT}=1/8$, while the order-parameter exponent $\beta_{\rm AT}$ and correlation-length exponent $\nu_{\rm AT}$ vary continuously, illustrating weak universality. Along the eight-vertex critical line of the XYZ chain, the critical exponents $\beta_{\rm 8v}$ and $\nu_{\rm 8v}$ as well as the scaling dimension $d_{\rm 8v}=\beta_{\rm 8v}/\nu_{\rm 8v}$ vary continuously, while the exponents satisfy the universal relation $4\beta_{\rm 8v}=2\nu_{\rm 8v}-1$ \cite{Baxter1972wg,Chepiga_2023}. These distinct scaling behaviors are consistent with the non-local duality: local order parameters in one description correspond to a topological order encoded in non-local string observables in the other.\\

At low energies, both critical lines are described by compact-boson $c=1$  Gaussian theories \cite{DiFrancesco1997,giamarchi2004quantum}, and specifically a $\mathbb{Z}_2$ orbifold theory for the AT model \cite{PhysRevB521138}. Remarkably, the eight-vertex transition in the XYZ chain also realizes one-dimensional deconfined quantum criticality, as it is a direct continuous transition between phases with incompatible spontaneously broken $\mathbb{Z}_2$ symmetries \cite{Senthil2004Deconfined,Pronk2025Deconfined}.\\

Despite their fundamental role in statistical mechanics and quantum many-body physics, experimental access to continuously varying criticality remains limited. In classical systems, signatures of Ashkin--Teller behavior have been observed in surface adsorbates \cite{PhysRevLett541539}. More recently, Rydberg-atom experiments have accessed related quantum critical behavior, including transitions associated with the melting of period-$4$ order and the vicinity of Ashkin--Teller critical points \cite{2019Natur.568..207K,Zhang2025Floating}. These experiments provide access to a specific realization of the Ashkin--Teller model, but have no mechanism to manipulate the details of the model to observe varying critical exponents of the Ashkin-Teller critical theory. Moreover, Ashkin--Teller criticality is realized only at a single fine-tuned point at the boundary of the period-$4$ phase, and with the presently available experimental resolution, accurately locating this critical point remains an additional challenge \cite{PhysRevB.110.125113}.\\
Complementing these experimental developments, theoretical proposals and numerical studies have shown how Ashkin--Teller criticality can emerge in Rydberg systems and, in multicomponent architectures, how the properties of the resulting multicritical point can be tuned \cite{Chepiga_2021,PhysRevLett132076505,PhysRevResearch7013215}. In these constructions, Ashkin--Teller criticality arises at fine-tuned conformal or multicritical points rather than as a directly accessible critical line with a control over the effective  Ashkin--Teller interaction. Likewise, to the best of our knowledge, eight-vertex criticality and its continuously varying critical exponents have not yet been experimentally demonstrated in a one-dimensional quantum system.\\

A natural route toward realizing interacting $c=1$ critical theories emerges from fermionic representations of spin models. Under the Jordan--Wigner transformation, each transverse-field Ising chain maps onto a Kitaev chain, allowing the quantum Ashkin--Teller model to be viewed as two Kitaev chains coupled through four-Majorana interactions. This makes recently developed quantum-dot Kitaev-chain architectures \cite{Sau_2012,Leijnse_2012,Fulga_2013,Liu_2022,Liu_2025} particularly promising building blocks: few-site chains have already been realized \cite{Dvir_2023,ten_Haaf_2024,Bordin_2025} using spin-polarized quantum dots connected through superconducting hybrid regions that generate elastic co-tunneling and crossed Andreev reflection, while gate control provides local tunability of the microscopic parameters. Extending these platforms to interacting critical models requires additional, model-dependent interaction mechanisms, which we develop below for the Ashkin--Teller and eight-vertex models.\\

The remainder of the paper is organized as follows. We first develop the theoretical framework and show how the proposed interactions generate the Ashkin--Teller and eight-vertex models, providing complementary routes to continuously varying quantum criticality. We then present concrete quantum-dot implementations and discuss how experimentally controllable parameters tune each system along its critical line, together with the practical limitations on the accessible interaction range.

\section{Theory}
\subsection{Ashkin--Teller model}
We begin by considering the quantum Ashkin--Teller model in its most common spin representation \cite{PhysRevB245229},
\begin{align}
H_{\mathrm{AT}}
&=\;
t_1\sum_{\alpha=1,2}\sum_n X_{\alpha,n}
+t_2\sum_{\alpha=1,2}\sum_n Z_{\alpha,n}Z_{\alpha,n+1}
\nonumber\\
&-\lambda_x\sum_n X_{1,n}X_{2,n}
-\lambda_z\sum_n
Z_{1,n}Z_{1,n+1}
Z_{2,n}Z_{2,n+1},
\label{eq:AT_spin}
\end{align}
where $X_{\alpha,n}$ and $Z_{\alpha,n}$ denote Pauli operators acting on site $n$ of chain $\alpha$.\\
To make a connection with an implementation in terms of fermions, as required for a quantum-dot realization, it is useful to express the model in terms
of Majorana fermions. We map each spin site onto a pair of Majorana operators, $a_{\alpha,n}$ and $b_{\alpha,n}$, with
\begin{align*}
c_{\alpha,n}
&=
\frac{1}{2}
\left(
a_{\alpha,n}
+i b_{\alpha,n}
\right),
&
c_{\alpha,n}^\dagger
&=
\frac{1}{2}
\left(
a_{\alpha,n}
-i b_{\alpha,n}
\right),
\end{align*}
and apply a Jordan--Wigner transformation independently to each chain,
\begin{align*}
X_{\alpha,n}
&= i a_{\alpha,n}b_{\alpha,n},\\
Z_{\alpha,n}Z_{\alpha,n+1}
&= i b_{\alpha,n}a_{\alpha,n+1}.
\end{align*}
Then, Eq.~\eqref{eq:AT_spin} takes the Majorana form \cite{Fabrizio_2000}
\begin{align}
H_M
&=\;
t_1\sum_{\alpha=1,2}\sum_n
i a_{\alpha,n}b_{\alpha,n}
+
t_2\sum_{\alpha=1,2}\sum_n
i b_{\alpha,n}a_{\alpha,n+1}
\nonumber\\
&+
\lambda_x\sum_n
a_{1,n}b_{1,n}a_{2,n}b_{2,n}
+
\lambda_z\sum_n
b_{1,n}a_{1,n+1}
b_{2,n}a_{2,n+1}.
\label{eq:AT_majorana}
\end{align}
Thus, the Ashkin--Teller model can equivalently be viewed as two coupled Kitaev
chains with local parity-preserving four-Majorana interactions. The
$\lambda_x$ and $\lambda_z$ terms couple, respectively, the on-site and
intersite Majorana bilinears of the two chains.\\

Since Majorana degrees of freedom are not directly accessible in experiment, we reformulate $H_{\rm M}$ in terms of Kitaev chains, whose microscopic degrees of freedom are spinless fermions \cite{Kitaev_2001}. For two independent chains, the Kitaev-chain Hamiltonian is
\begin{align}
H_{{\rm K},\alpha}
&=
-\sum_j
\mu_{\alpha,j}
\left(
n_{\alpha,j}-\frac12
\right)
-\sum_j
t_{\alpha,j}
\left(
c_{\alpha,j}^\dagger c_{\alpha,j+1}
+\text{h.c.}
\right)
\nonumber\\
&\quad
+
\sum_j
\Delta_{\alpha,j}
\left(
c_{\alpha,j}c_{\alpha,j+1}
+\text{h.c.}
\right),
\end{align}
where $\mu_{\alpha,j}$ denotes the local chemical potential, $t_{\alpha,j}$ the nearest-neighbor hopping amplitude, $\Delta_{\alpha,j}$ the superconducting pairing amplitude, and the number operator is defined as
\[
n_{\alpha,j}\equiv c_{\alpha,j}^\dagger c_{\alpha,j}.
\]

At the special point
\[
t_{\alpha,j}=\Delta_{\alpha,j},
\qquad \forall\,\alpha,j,
\]
the Kitaev-chain Hamiltonian can be written purely in terms of nearest-neighbor Majorana bilinears and becomes equivalent to the Majorana Hamiltonian
\[
H_{\rm M}(\lambda_x=\lambda_z=0).
\]
To engineer the Ashkin--Teller interactions in the Kitaev chain, we introduce two types of resonator modes. First, we consider a local bosonic mode associated with site $j$, with dimensionless displacement operator
\begin{equation*}
\hat Q_{x,j}
=
a_{x,j}
+
a_{x,j}^\dagger,
\end{equation*}
where $a_{x,j}$ ($a_{x,j}^\dagger$) annihilates (creates) an excitation of the corresponding resonator mode. This mode couples capacitively to the quantum dots and modulates the local chemical potentials of both chains according to
\begin{equation}
\mu_{\alpha,j}
\rightarrow
\mu_{\alpha,j}
+
\eta_{\alpha,j}^{(x)}
\hat Q_{x,j}.
\label{eq mu variation}
\end{equation}
Here, $\eta_{\alpha,j}^{(x)}$ denotes the coupling strength between the site resonator mode and the chemical potential of chain $\alpha$.
The corresponding coupling to the onsite charge degrees of freedom is
\begin{align*}
H_x
&\equiv
-\sum_{\alpha,j}
\eta_{\alpha,j}^{(x)}
\hat Q_{x,j}
\left(
n_{\alpha,j}-\frac12
\right)
\nonumber\\
&=
-\sum_{\alpha,j}
\eta_{\alpha,j}^{(x)}
\hat Q_{x,j}
\frac{i}{2}
a_{\alpha,j}b_{\alpha,j}.
\end{align*}
Introducing the on-site parity operator
\begin{equation*}
P_{\alpha,j}^{x}
\equiv
i a_{\alpha,j}b_{\alpha,j}
=
2n_{\alpha,j}-1,
\end{equation*}
this can be written as
\begin{equation*}
H_x
=
\sum_j
\hat Q_{x,j}
\left[
g^x_{1,j}P_{1,j}^{x}
+
g^x_{2,j}P_{2,j}^{x}
\right],
\end{equation*}
where the constants \(g^x_{\alpha,j}\) absorb the factors
\(-\eta_{\alpha,j}^{(x)}/2\).

Second, a link mode
\begin{equation*}
\hat Q_{z,j}
=
a_{z,j}
+
a_{z,j}^\dagger
\end{equation*}
modulates both the hopping and induced pairing on the link \((j,j+1)\):
\begin{equation}
t_{\alpha,j}
\rightarrow
t_{\alpha,j}
+
\eta_{\alpha,j}^{(t)}
\hat Q_{z,j},
\qquad
\Delta_{\alpha,j}
\rightarrow
\Delta_{\alpha,j}
+
\eta_{\alpha,j}^{(\Delta)}
\hat Q_{z,j}.
\label{eq t del variation}
\end{equation}
The resonator-dependent link Hamiltonian is
\begin{align*}
H_z
&\equiv
\sum_{\alpha,j}
\frac{i\hat Q_{z,j}}{2}
\left[
\left(
\eta_{\alpha,j}^{(t)}
+
\eta_{\alpha,j}^{(\Delta)}
\right)
b_{\alpha,j}a_{\alpha,j+1}\right.\\
&+\left.
\left(
\eta_{\alpha,j}^{(\Delta)}
-
\eta_{\alpha,j}^{(t)}
\right)
a_{\alpha,j}b_{\alpha,j+1}
\right]\\
&\equiv
\sum_j
\hat Q_{z,j}
\left[
g^z_{1,j}P_{1,j}^{z}
+
g^z_{2,j}P_{2,j}^{z}
\right],
\end{align*}
where we choose $\eta_{\alpha,j}^{(t)}
=
\eta_{\alpha,j}^{(\Delta)}$, such that
the unwanted \(i a_{\alpha,j}b_{\alpha,j+1}\) coupling vanishes.\\

The bosonic resonator Hamiltonian is
\begin{equation*}
H_{\rm res}
=
\sum_j
\omega_{x,j} a_{x,j}^\dagger a_{x,j}
+
\sum_j
\omega_{z,j} a_{z,j}^\dagger a_{z,j}.
\end{equation*}
Here the resonator frequencies are allowed to depend on position. The
anti-adiabatic regime is then defined by
\begin{equation*}
\omega_{x,j},\omega_{z,j}
\gg
E_{\rm Kitaev}
\qquad
\text{for all } j .
\end{equation*}
In this limit the resonator modes can be eliminated perturbatively, for
example by a Schrieffer--Wolff transformation \cite{Bravyi2011}. The second-order effective Hamiltonian is therefore
\begin{align*}
H_{\rm eff}&\equiv H_{\rm eff}^{x}+H_{\rm eff}^{z}\\
&=
{\rm const.}
-
\sum_j
\frac{2g^x_{1,j}g^x_{2,j}}{\omega_{x,j}}
P_{1,j}^{x}P_{2,j}^{x}\\
&-
\sum_j
\frac{2g^z_{1,j}g^z_{2,j}}{\omega_{z,j}}
P_{1,j}^{z}P_{2,j}^{z},
\end{align*}
which corresponds to the interaction term of the Ashkin--Teller model. From the expression for $H_{\rm eff}$ above, we see that both
$\lambda_{x,j}$ and $\lambda_{z,j}$ are determined by the coupling
between the Kitaev chains and the resonator, as well as by the resonator
frequency.\\

To recover a clean translation invariant Ashkin--Teller chain one requires
\begin{equation*}
\lambda_{x,j}= \lambda_x,
\qquad
\lambda_{z,j}= \lambda_z, \qquad \forall\,j.
\end{equation*}
In the symmetric self-dual case,
\begin{equation*}
    \lambda_x=\lambda_z\equiv \lambda,
    \qquad
    t_1=t_2\equiv t,
\end{equation*}
the model lies on the Ashkin--Teller critical line. Therefore, varying
\(\lambda/t\) while maintaining the self-dual condition provides a direct
way to tune continuously along the critical line, changing the critical
exponents without leaving the gapless Ashkin--Teller universality class.\\ 

\subsection{Eight-vertex model}
The second model we consider is the XYZ model:

\begin{equation}
H_{\rm XYZ}
=
\sum_j
\left[
J_x X_jX_{j+1}
+
J_yY_jY_{j+1}
+
J_zZ_jZ_{j+1}
\right].
\label{eq XYZ}
\end{equation}

In terms of Dirac fermions,
the Hamiltonian becomes \cite{Chepiga_2023}
\begin{equation}
\begin{aligned}
H
&=
-t\sum_j
\left(
c_j^\dagger c_{j+1}
+
c_{j+1}^\dagger c_j
\right)
+
\Delta\sum_j
\left(
c_jc_{j+1}
+
c_{j+1}^\dagger c_j^\dagger
\right)
\\
&\quad
+
V\sum_j
\left(
n_j-\frac12
\right)
\left(
n_{j+1}-\frac12
\right),
\end{aligned}
\label{eq XYZ kitaev}
\end{equation}
with
\[
t=J_x+J_y,
\qquad
\Delta=J_x-J_y,
\qquad
V=4J_z.
\]
The interaction term $V$ in the above Hamiltonian corresponds to a nearest-neighbor Coulomb interaction in the Kitaev chain, of the form $n_j n_{j+1}$. It additionally induces a shift of the effective chemical potential.
Eight-vertex criticality corresponds to the condition $J_x=-J_z$ \cite{PhysRevLett26834,Chepiga_2023}, which in the parametrization above translates to $t+\Delta=-V/2$. Analogous to tuning the interaction parameter $\lambda$ along the Ashkin--Teller critical line, here one can move continuously along the eight-vertex critical line by varying $J_y=(t-\Delta)/2$ while keeping $J_x=(t+\Delta)/2$ fixed.\\

The correspondence between observables in the AT and eight-vertex models can be expressed using
the AT disorder operators
\[
\mu_{\sigma,j-1/2}=\prod_{\ell<j}\sigma_\ell^x,
\qquad
\mu_{\tau,j-1/2}=\prod_{\ell<j}\tau_\ell^x.
\]
For open chains and $j<k$, the transformation yields
\cite{PhysRevA81032334,FENDLEY1989549}
\begin{align*}
&\langle Z_{2j-1}Z_{2k-1}\rangle_{\mathrm{XYZ}}
\\
&\quad =
\bigl\langle
(\sigma_j^z\mu_{\tau,j-1/2})
(\sigma_k^z\mu_{\tau,k-1/2})
\bigr\rangle_{\mathrm{AT}},
\\[4pt]
&\langle X_{2j-1}X_{2k-1}\rangle_{\mathrm{XYZ}}
\\
&\quad =
\bigl\langle
(\tau_j^z\mu_{\sigma,j-1/2})
(\tau_k^z\mu_{\sigma,k-1/2})
\bigr\rangle_{\mathrm{AT}},
\end{align*}
where $\sigma_i^{x,y,z}$ and $\tau_i^{x,y,z}$ denote the Pauli operators acting on site $i$ of chains 1 and 2, respectively. This form implies a string of Pauli matrices between sites $j$ and $k$.
Local XYZ ordering therefore probes AT order--disorder
composites, whose scaling dimensions generally differ from
those of the local AT magnetization and polarization.

\section{Proposal for experimental realization in quantum dot arrays}
In this section, we explain how quantum-dot arrays can realize the Ashkin--Teller and eight-vertex models through resonator-mediated and Coulomb interactions, respectively.
\subsection{Ashkin--Teller model}
In the quantum-dot implementation, the resonator-induced modulation of the chemical potential introduced in Eq.~\eqref{eq mu variation} can be realized \cite{Harvey_2018} by capacitively coupling the plunger gates of the two quantum dots to a common resonator. Similarly, the modulation of the hopping and superconducting pairing amplitudes in Eq.~\eqref{eq t del variation} can be generated by coupling the resonator to the left or right tunnel gates of the Andreev bound states connecting neighboring dots in the two chains.

\begin{figure}[h]
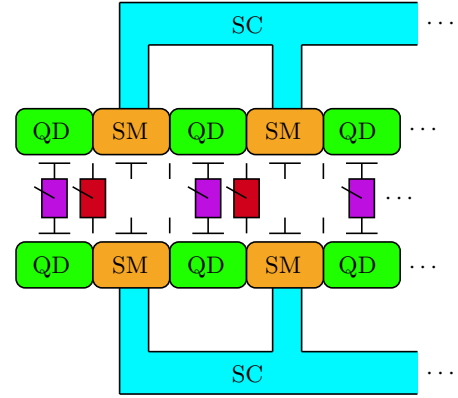

    \centering
    \include{expsetup_rescaled}
    \caption{Experimental setup to realize the Ashkin--Teller model using two Kitaev chains coupled to resonators (purple and red represent $x$ and $z$ modulation respectively). The Kitaev chain is formed by an alternating array of quantum dots (QDs) and proximitized semiconductor segments hosting Andreev bound states (SM+SC).}
    \label{fig exp system}
\end{figure}

Let $t_L$ and $t_R$ denote the tunnel couplings between the quantum dots and the intermediate Andreev bound state. The resulting effective elastic-cotunneling and crossed-Andreev-reflection amplitudes scale \cite{Liu_2022} as

\begin{align*}
    t &\sim t_Lt_R f_1(\theta)\frac{\mu_{ABS}}{\mu_{ABS}^2+\Delta_{ABS}^2}\\
    \Delta &\sim t_Lt_R f_2(\theta)\frac{\Delta_{ABS}}{\mu_{ABS}^2+\Delta_{ABS}^2},
\end{align*}

where $\mu_{\mathrm{ABS}}$ and $\Delta_{\mathrm{ABS}}$ characterize the Andreev bound state. The functions $f_1(\theta)$ and $f_2(\theta)$ encode the dependence of elastic co-tunneling and crossed Andreev reflection on the relative orientation of the spin--orbit field and the applied magnetic field. Since both amplitudes depend on the product $t_Lt_R$, coupling the resonator to either tunnel gate produces the required simultaneous modulation of the effective hopping and pairing amplitudes. The corresponding experimental situation is illustrated in Fig.~\ref{fig exp system}.

For representative Kitaev-chain parameters $t=\Delta=10~\mu\mathrm{eV}$, we consider resonator frequencies $\omega_r/2\pi=10$--$12~\mathrm{GHz}$ and dot--resonator couplings $g/2\pi=1$--$1.5~\mathrm{GHz}$ \cite{Bordin_2025}. These values should be regarded as experimentally demanding, in particular the upper end of the coupling range, but they provide a useful estimate of the interaction strengths that could be targeted in an optimized device. In the anti-adiabatic regime, integrating out the resonator generates an effective interaction of order \[ \lambda_{\mathrm{eff}}\simeq \frac{2g^2}{\omega_r}. \] Using $h\times1~\mathrm{GHz}\simeq4.14~\mu\mathrm{eV}$, a coupling $g/2\pi=1~\mathrm{GHz}$ gives $\lambda_{\mathrm{eff}}\simeq0.83~\mu\mathrm{eV}$ for $\omega_r/2\pi=10~\mathrm{GHz}$ and $\lambda_{\mathrm{eff}}\simeq0.69~\mu\mathrm{eV}$ for $12~\mathrm{GHz}$. Increasing the coupling to $g/2\pi=1.5~\mathrm{GHz}$ yields approximately $1.86~\mu\mathrm{eV}$ and $1.55~\mu\mathrm{eV}$, respectively. For $t=\Delta=10~\mu\mathrm{eV}$, this corresponds to \[ \frac{\lambda_{\mathrm{eff}}}{t}\simeq0.07-0.19. \]

Thus, resonator parameters in this regime enable interaction strengths at the $10–20\%$ level of the bare Kitaev-chain energy scale, providing direct access to a finite portion of the Ashkin--Teller critical line, ranging approximately from $\lambda_{\mathrm{eff}}/t\approx0$ to $\lambda_{\mathrm{eff}}/t\approx0.19$. Further advances in resonator coupling would naturally extend this accessible range toward progressively stronger interactions. Crosstalk between neighboring resonators can be suppressed by slightly staggering their resonance frequencies through variations of the inductance. Since the effective interaction scales as $\lambda \propto \alpha^2/C$, this leaves the interaction strength unchanged as long as the capacitance $C$ and resonator--dot lever arm $\alpha$ remain fixed.\\

The continuously varying critical exponents of the Ashkin--Teller model can
be extracted from finite-size scaling of the magnetic and polarization order
parameters \cite{PhysRevB76224423}. For an open chain with fixed boundary conditions, we define the
center observables
\[
m_Z(L)
\equiv
\left\langle Z_{\alpha,L/2}\right\rangle_{\rm fixed},
\qquad
m_P(L)
\equiv
\left\langle Z_{1,L/2}Z_{2,L/2}\right\rangle_{\rm fixed}.
\]
In the dual representation relevant to the quantum-dot implementation, these
operators are represented by the parity strings
\[
S_Z^i(L)
=
\left\langle
\prod_{j=0}^{L/2-1} X_{i,j}
\right\rangle,
\qquad
S_P(L)
=
\left\langle
\prod_{j=0}^{L/2-1} X_{1,j}X_{2,j}
\right\rangle.
\]
Since these dual observables correspond to the same critical scaling
operators, they are governed by the same scaling dimensions $x_Z$ and $x_P$.

For a finite lattice, the boundary-CFT scaling form \cite{Cardy2006} that we will use in our numerical
analysis is
\begin{equation}
\begin{aligned}
\left|S_Z(L)\right|
&\equiv
\left|
\frac{S_Z^1(L)+S_Z^2(L)}{2}
\right|
\\
&\propto
\left|
(L+1)
\sin\left[
\frac{\pi L}{2(L+1)}
\right]
\right|^{-x_Z}.
\end{aligned}
\label{eq scaling1}
\end{equation}
and
\begin{equation}
\left|S_P(L)\right|
\propto
\left|
(L+1)
\sin\left[
\frac{\pi L}{2(L+1)}
\right]
\right|^{-x_P}.
\label{eq scaling2}
\end{equation}
Here, $S_Z(L)$ is obtained by averaging the parity strings of the two chains, which are equivalent by the $\mathbb{Z}_2$ symmetry of the symmetric Ashkin--Teller model.\\

In the large-$L$ limit, the conformal distance scales as $L$, recovering the
asymptotic behavior
\[
|m_Z(L)|\sim |S_Z(L)|\propto L^{-x_Z},
\quad
|m_P(L)|\sim |S_P(L)|\propto L^{-x_P}.
\]
Since $X_{\alpha,j}$ corresponds to the local fermion parity $P_{\alpha,j}^{x}$, these observables can be reconstructed experimentally from repeated local charge-readout measurements. In practice, the state can first be frozen by rapidly isolating the quantum dots, after which their occupations are measured. Repeating this procedure yields the parity-string expectation values by averaging the corresponding products of measured parities over many experimental realizations.\\

\begin{figure}[h]
    \centering
\includegraphics[width=\linewidth]{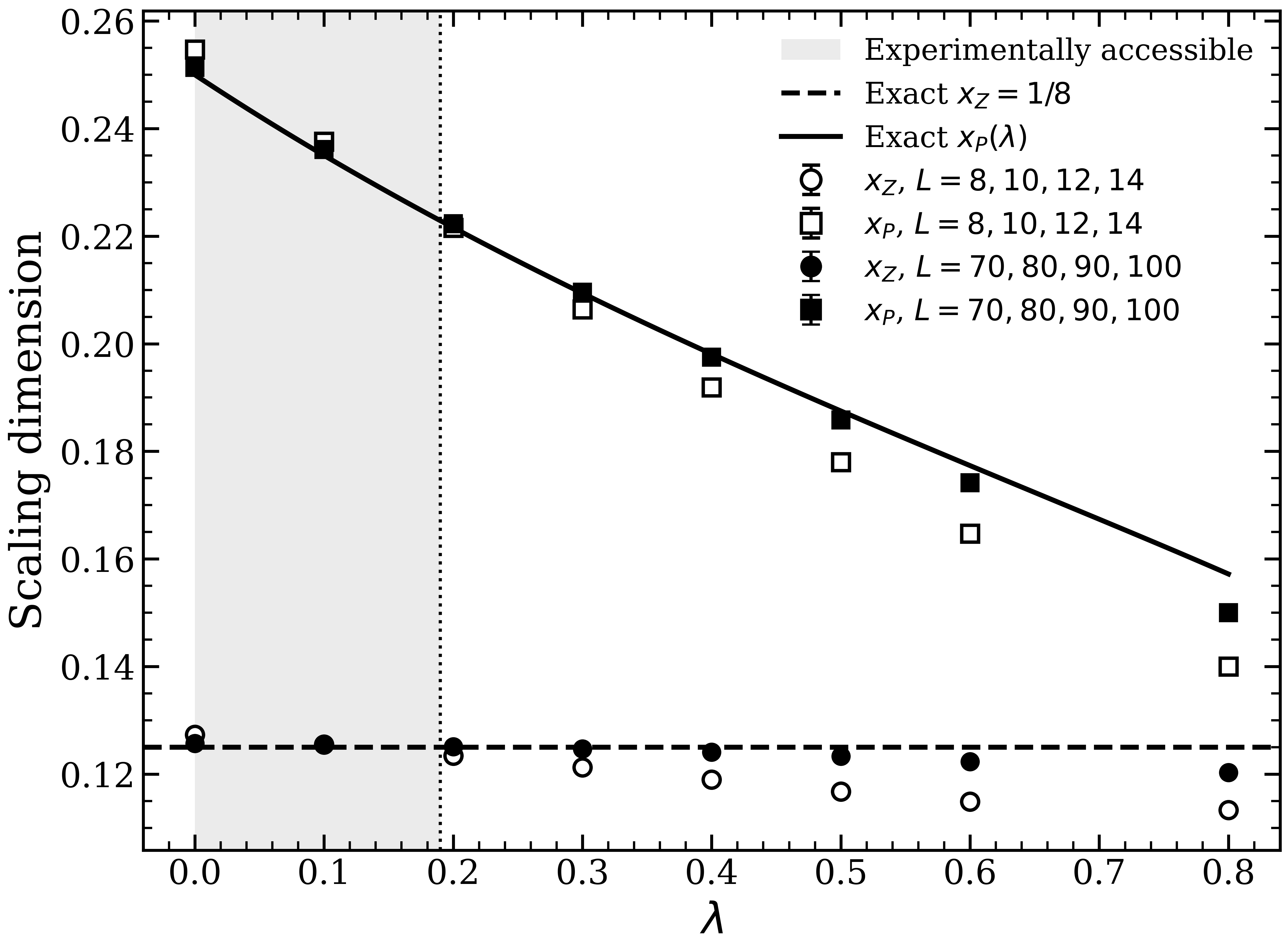}
    \caption{The magnetic and polarization scaling dimensions, $x_m$ and $x_P$, along the Ashkin--Teller critical line, obtained using finite-size scaling. Symbols show the scaling dimensions extracted from DMRG calculations, while the solid and dashed lines denote the corresponding exact critical-theory predictions. The shaded region indicates the experimentally accessible range $0\leq\lambda\leq0.19$. The fit data underlying this figure are provided in Appendix~\ref{app:fss}.}
    \label{figATrange}
\end{figure}

Figure~\ref{figATrange} shows density-matrix renormalization-group (DMRG) \cite{White1992,OstlundRommer1995,Schollwock2011} 
calculations of the scaling
dimensions $x_P$ and $x_Z$ for the small system sizes
$L=8,10,12,14$. The scaling dimensions are extracted from fits to the logarithmic forms of
the boundary-CFT predictions given in Eqs.~\eqref{eq scaling1} and
\eqref{eq scaling2}. Along the Ashkin--Teller critical line, the exact  \cite{PhysRevB76224423}
scaling dimensions are 
\[
x_Z=\frac{1}{8},
\qquad
x_P(\lambda)=
\frac{\pi}{8\arccos(-\lambda)}.
\]

The AT polarization is an order--order product, whereas the XYZ magnetic order parameter on the eight-vertex critical line maps to an AT order--disorder product. At corresponding critical points, their scaling dimensions are inversely related through
$x_{\mathrm{XYZ}}=1/(16x_P)$
\cite{Delfino_2004}.\\

Although $x_Z$ is independent of $\lambda$ in the thermodynamic limit, the
finite-size estimates display a systematic $\lambda$-dependent drift due to finite-size effects. 
These corrections become increasingly pronounced as
the four-state Potts point is approached, where logarithmic corrections are
particularly strong \cite{PhysRevB91165129}. Consequently, the drift is expected to decrease when
larger system sizes are used in the finite-size scaling analysis, as confirmed
by the large-$L$ results shown in Fig.~\ref{figATrange}. Importantly, within
the experimentally accessible range of $\lambda$, the ratio of the scaling
dimensions, $x_P(\lambda)$ changes by approximately $10\%$, providing a sizable experimentally
resolvable signature of the continuously varying critical behavior.
\\
\subsection{Eight-vertex model}
As discussed in the previous section, eight-vertex criticality is obtained within a particular parameter regime of the XYZ model. Here, we propose a quantum-dot implementation in which the required interaction is generated directly by the Coulomb coupling between neighboring dots. In the fermionic representation, this capacitive interaction gives rise to the nearest-neighbor density-density term entering the XYZ Hamiltonian. In a realistic device, additional Coulomb contributions, such as longer-range intrachain interactions, may also be present. These unwanted terms can be mitigated through suitable device geometry and electrostatic screening by nearby metallic gates. The corresponding experimental setup is sketched in Fig.~\ref{fig exp system 2}.\\

A key advantage of this implementation is that the interaction is generated directly by the Coulomb coupling between neighboring dots, rather than through a resonator-mediated process. Consequently, the interaction strength is not limited by the experimentally achievable dot--resonator coupling or resonator frequency and can be made comparable to, or even larger than, the hopping and pairing scales. In closely related quantum-dot Kitaev-chain devices, nonlocal charging energies of approximately $20$--$30~\mu\mathrm{eV}$ have already been extracted \cite{bordin2025probingmajoranalocalizationphasecontrolled}, compared with characteristic hopping and pairing amplitudes of order $t,\Delta\sim10~\mu\mathrm{eV}$, corresponding to interaction strengths several times the single-particle energy scale. Moreover, substantially larger capacitive couplings have been demonstrated in other quantum-dot geometries \cite{PhysRevApplied12064049}, indicating that interactions of several tens of $\mu\mathrm{eV}$ should be accessible through suitable device design and electrostatic screening. This provides access to a broad range of the XYZ phase diagram, including the eight-vertex critical line.\\

For the XYZ model, the situation is particularly favorable because the local density directly corresponds to the order parameter. Consequently, measuring the local charge density provides direct access to the local order parameter, without requiring the evaluation of a nonlocal parity string. \\

\begin{figure}[h]
    \centering
\includegraphics[width=\linewidth]{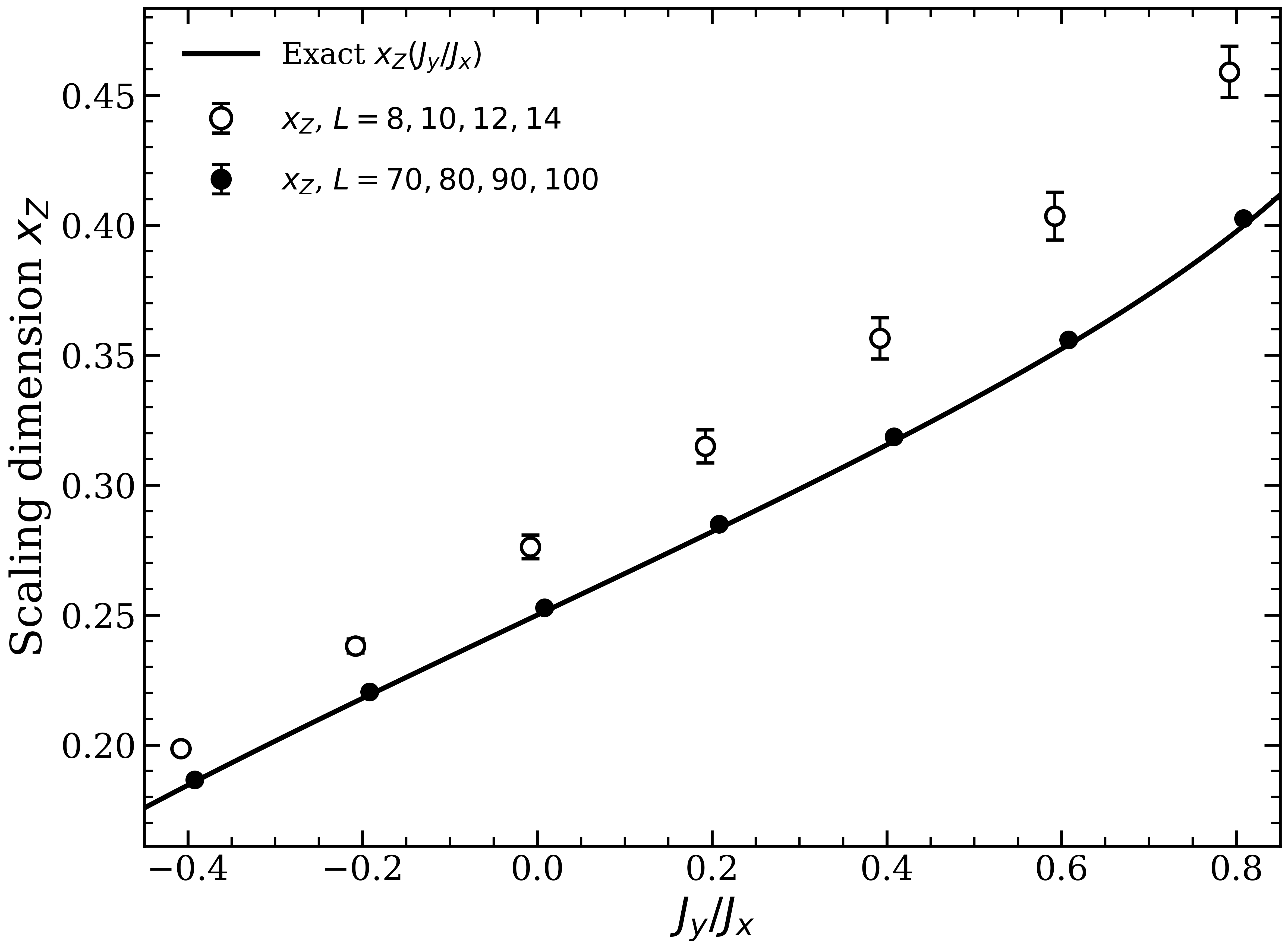}
    \caption{The magnetic scaling dimension $x_m$ along the
eight-vertex critical line obtained using finite-size scaling. Symbols show the scaling dimensions extracted
from DMRG calculations, while the solid line denotes the
exact eight-vertex prediction. The fit data underlying this figure are provided in Appendix~\ref{app:fss}.}
    \label{figXYZrange}
\end{figure}

Figure~\ref{figXYZrange} compares the magnetic scaling dimension $x_Z$
extracted from finite-size DMRG calculations with the exact eight-vertex
prediction as a function of $J_y/J_x$. Along the eight-vertex critical line, the exact magnetic
scaling dimension is \cite{Baxter1972wg,Chepiga_2023}
\[
x_Z(J_y/J_x)=\frac{1}{2}-\frac{1}{2\pi}\arccos{(J_y/J_x)}.
\] Following Ref.~\cite{Chepiga_2023},
the scaling dimension is extracted from the finite-size scaling of the
Friedel-oscillation amplitude near the center of the chain,
\[
D(L)\equiv
\left|n_{L/2-1}-n_{L/2}\right|
\propto
\left|
(L+1)
\sin\left[
\frac{\pi L}{2(L+1)}
\right]
\right|^{-x_Z},
\]
where evaluating the oscillation amplitude close to the chain center
minimizes boundary-induced corrections.
For experimentally accessible Coulomb interaction strengths, the magnetic
scaling dimension can exhibit substantially larger relative variations than
the approximately $10\%$ changes expected in the polarization scaling
dimension of the Ashkin--Teller realization. We compare fits obtained from relatively small
system sizes with those based on larger systems. The residual drift with
$J_y/J_x$ is attributable to finite-size corrections and is substantially
reduced as the system size is increased.

The same idea can also be applied to a two-chain geometry. In this case, the Coulomb interaction between quantum dots located on the same rung directly generates the interchain density-density coupling required for the asymmetric Ashkin--Teller model with only the $\lambda_x$ interaction present. As in the eight-vertex implementation, this direct electrostatic coupling allows access to interaction strengths that are not restricted by the resonator-mediated mechanism.
\begin{figure}[h]
    \centering
    \vspace{5mm}
    \tikzset{every picture/.style={line width=0.6pt}} 

\begin{tikzpicture}[x=0.6pt,y=0.6pt,yscale=-1,xscale=1]
\draw [color={rgb, 255:red, 0; green, 0; blue, 0 }  ,draw opacity=1 ][line width=1.3]    (234.2,1481.82) -- (208.2,1512.41) ;
\draw [color={rgb, 255:red, 208; green, 2; blue, 27 }  ,draw opacity=1 ]   (39.5,1525.57) .. controls (103.77,1560.95) and (168.03,1560.95) .. (232.3,1525.57) ;
\draw  [color={rgb, 255:red, 208; green, 2; blue, 27 }  ,draw opacity=1 ] (232.3,1525.57) .. controls (168.03,1490.19) and (103.77,1490.19) .. (39.5,1525.57) ;
\draw  [draw opacity=0][fill={rgb, 255:red, 0; green, 248; blue, 255 }  ,fill opacity=1 ] (80.36,1449.37) -- (98.56,1449.37) -- (98.56,1489.37) -- (80.36,1489.37) -- cycle ;
\draw  [draw opacity=0][fill={rgb, 255:red, 0; green, 248; blue, 255 }  ,fill opacity=1 ] (176.36,1449.37) -- (194.56,1449.37) -- (194.56,1489.37) -- (176.36,1489.37) -- cycle ;
\draw  [draw opacity=0][fill={rgb, 255:red, 0; green, 248; blue, 255 }  ,fill opacity=1 ] (80.36,1422.77) -- (267.2,1422.77) -- (267.2,1449.37) -- (80.36,1449.37) -- cycle ;
\draw [color={rgb, 255:red, 0; green, 0; blue, 0 }  ,draw opacity=1 ][line width=1.3]    (137.8,1481.82) -- (111.8,1512.41) ;
\draw  [fill={rgb, 255:red, 245; green, 166; blue, 35 }  ,fill opacity=1 ] (63.6,1495.13) .. controls (63.6,1491.95) and (66.18,1489.37) .. (69.36,1489.37) -- (106.04,1489.37) .. controls (109.22,1489.37) and (111.8,1491.95) .. (111.8,1495.13) -- (111.8,1512.41) .. controls (111.8,1515.59) and (109.22,1518.17) .. (106.04,1518.17) -- (69.36,1518.17) .. controls (66.18,1518.17) and (63.6,1515.59) .. (63.6,1512.41) -- cycle ;
\draw  [fill={rgb, 255:red, 245; green, 166; blue, 35 }  ,fill opacity=1 ] (160,1495.13) .. controls (160,1491.95) and (162.58,1489.37) .. (165.76,1489.37) -- (202.44,1489.37) .. controls (205.62,1489.37) and (208.2,1491.95) .. (208.2,1495.13) -- (208.2,1512.41) .. controls (208.2,1515.59) and (205.62,1518.17) .. (202.44,1518.17) -- (165.76,1518.17) .. controls (162.58,1518.17) and (160,1515.59) .. (160,1512.41) -- cycle ;
\draw    (80.36,1422.77) -- (80.36,1489.37) ;
\draw    (80.36,1422.77) -- (267.2,1422.77) ;
\draw    (98.56,1449.37) -- (176.36,1449.37) ;
\draw    (98.56,1489.37) -- (98.56,1449.37) ;
\draw    (176.36,1489.37) -- (176.36,1449.37) ;
\draw    (194.56,1489.37) -- (194.56,1449.37) ;
\draw    (267.2,1449.37) -- (194.56,1449.37) ;
\draw  [line width=1.3]   (29.6,1544.56) -- (49.4,1544.56) ;
\draw [line width=1.3][color={rgb, 255:red, 0; green, 0; blue, 0 }  ,draw opacity=0.45 ]   (80.36,1514.9) -- (100.16,1514.9) ;
\draw  [line width=1.3]  (126,1544.56) -- (145.8,1544.56) ;
\draw [line width=1.3] [color={rgb, 255:red, 0; green, 0; blue, 0 }  ,draw opacity=0.45 ]   (176.36,1514.9) -- (196.16,1514.9) ;
\draw  [line width=1.3]  (222.4,1544.56) -- (242.2,1544.56) ;
\draw  [line width=1.3]  (39.5,1544.56) -- (26.5,1559.85) ;
\draw  [line width=1.3]  (135.9,1544.56) -- (122.9,1559.85) ;
\draw  [line width=1.3]  (232.3,1544.56) -- (219.3,1559.85) ;
\draw [line width=1.3][color={rgb, 255:red, 0; green, 0; blue, 0 }  ,draw opacity=0.45 ]   (115.46,1484.31) -- (89.46,1514.9) ;
\draw [line width=1.3][color={rgb, 255:red, 0; green, 0; blue, 0 }  ,draw opacity=0.45 ]   (211.46,1484.31) -- (185.46,1514.9) ;
\draw [line width=1.3] [color={rgb, 255:red, 0; green, 0; blue, 0 }  ,draw opacity=0.45 ]   (89.6,1481.82) -- (63.6,1512.41) ;

\draw [line width=1.3] [color={rgb, 255:red, 0; green, 0; blue, 0 }  ,draw opacity=0.45 ]   (186,1481.82) -- (160,1512.41) ;
\draw  [color={rgb, 255:red, 208; green, 2; blue, 27 }  ,draw opacity=1 ]  (39.5,1525.57) -- (135.9,1525.57) ;
\draw [color={rgb, 255:red, 208; green, 2; blue, 27 }  ,draw opacity=1 ]   (135.9,1525.57) -- (232.3,1525.57) ;
\draw  [color={rgb, 255:red, 208; green, 2; blue, 27 }  ,draw opacity=1 ] (39.5,1525.57) .. controls (71.63,1534.28) and (103.77,1534.28) .. (135.9,1525.57) ;
\draw [color={rgb, 255:red, 208; green, 2; blue, 27 }  ,draw opacity=1 ]  (135.9,1525.57) .. controls (103.77,1518.28) and (71.63,1518.28) .. (39.5,1525.57) ;
\draw  [color={rgb, 255:red, 208; green, 2; blue, 27 }  ,draw opacity=1 ] (232.3,1525.57) .. controls (200.17,1518.28) and (168.03,1518.28) .. (135.9,1525.57) ;
\draw [color={rgb, 255:red, 208; green, 2; blue, 27 }  ,draw opacity=1 ]  (135.9,1525.57) .. controls (168.03,1534.28) and (200.17,1534.28) .. (232.3,1525.57) ;
\draw  [fill={rgb, 255:red, 31; green, 255; blue, 0 }  ,fill opacity=1 ] (111.8,1516.93) .. controls (111.8,1513.75) and (114.38,1511.17) .. (117.56,1511.17) -- (154.24,1511.17) .. controls (157.42,1511.17) and (160,1513.75) .. (160,1516.93) -- (160,1534.21) .. controls (160,1537.39) and (157.42,1539.97) .. (154.24,1539.97) -- (117.56,1539.97) .. controls (114.38,1539.97) and (111.8,1537.39) .. (111.8,1534.21) -- cycle ;
\draw  [color={rgb, 255:red, 208; green, 2; blue, 27 }  ,draw opacity=1 ][dash pattern={on 4.5pt off 4.5pt}] (232.3,1525.57) .. controls (168.03,1490.19) and (103.77,1490.19) .. (39.5,1525.57) ;
\draw  [fill={rgb, 255:red, 31; green, 255; blue, 0 }  ,fill opacity=1 ] (15.4,1516.93) .. controls (15.4,1513.75) and (17.98,1511.17) .. (21.16,1511.17) -- (57.84,1511.17) .. controls (61.02,1511.17) and (63.6,1513.75) .. (63.6,1516.93) -- (63.6,1534.21) .. controls (63.6,1537.39) and (61.02,1539.97) .. (57.84,1539.97) -- (21.16,1539.97) .. controls (17.98,1539.97) and (15.4,1537.39) .. (15.4,1534.21) -- cycle ;

\draw  [fill={rgb, 255:red, 31; green, 255; blue, 0 }  ,fill opacity=1 ] (208.2,1516.93) .. controls (208.2,1513.75) and (210.78,1511.17) .. (213.96,1511.17) -- (250.64,1511.17) .. controls (253.82,1511.17) and (256.4,1513.75) .. (256.4,1516.93) -- (256.4,1534.21) .. controls (256.4,1537.39) and (253.82,1539.97) .. (250.64,1539.97) -- (213.96,1539.97) .. controls (210.78,1539.97) and (208.2,1537.39) .. (208.2,1534.21) -- cycle ;

\draw  [color={rgb, 255:red, 208; green, 2; blue, 27 }  ,draw opacity=1 ] (131.31,1547.42) -- (141.19,1557.31)(141.44,1547.17) -- (131.06,1557.56) ;
\draw  [color={rgb, 255:red, 208; green, 2; blue, 27 }  ,draw opacity=1 ] (131.31,1494.42) -- (141.19,1504.31)(141.44,1494.17) -- (131.06,1504.56) ;

\draw (122,1518) node [anchor=north west][inner sep=0.75pt]   [align=left] {QD};
\draw (76,1495) node [anchor=north west][inner sep=0.75pt]   [align=left] {SM};
\draw (172,1495) node [anchor=north west][inner sep=0.75pt]   [align=left] {SM};
\draw (147,1428) node [anchor=north west][inner sep=0.75pt]   [align=left] {SC};
\draw (269,1428) node [anchor=north west][inner sep=0.75pt]   [align=left] {$\displaystyle \cdots $};
\draw (262,1510) node [anchor=north west][inner sep=0.75pt]   [align=left] {$\displaystyle \cdots $};
\draw (218,1518) node [anchor=north west][inner sep=0.75pt]   [align=left] {QD};
\draw (26,1518) node [anchor=north west][inner sep=0.75pt]   [align=left] {QD};
\end{tikzpicture}
    \caption{Experimental setup for realizing the eight-vertex model through Coulomb interactions between neighboring quantum dots. Longer-range Coulomb interactions (red lines) are screened (red crosses) by the metallic leads (black), effectively restricting the interaction to nearest neighbors. The Kitaev chain is formed by an alternating array of quantum dots (QDs) and 
    proximitized semiconductor segments hosting Andreev bound states (SM+SC).}
    \label{fig exp system 2}
\end{figure}
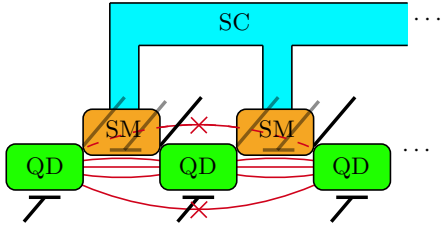\\

We end this section by noting that the proposed setup allows for several complementary routes to extract critical exponents and scaling dimensions. In addition to the finite-size scaling analyses considered above, the quantum Fisher information provides an alternative probe, since its universal scaling near criticality is directly governed by the scaling dimension of the corresponding collective observable and can be accessed experimentally through the dynamical susceptibility \cite{Hauke2016}.

\section{Conclusion}
In this work, we proposed quantum-dot architectures for realizing interacting
one-dimensional models with continuously varying $c=1$ criticality. We focused
on the Ashkin--Teller and eight-vertex models, which provide two complementary
settings in which critical exponents vary continuously along a quantum
critical manifold. Accessing such behavior experimentally is particularly
interesting because it goes beyond the more conventional situation in which
critical exponents are fixed solely by the symmetry of the ordered phase, and provides a
direct setting in which weak universality and continuously varying scaling
dimensions can be tested in a controlled microscopic system.\\

Resonator-mediated interactions provide a route to the quantum
Ashkin--Teller model, whereas direct Coulomb interactions enable the
realization of the XYZ model and its eight-vertex critical line. Combined
with the high degree of electrostatic control available in quantum-dot
arrays, these constructions allow the microscopic parameters governing the
critical theory to be tuned directly and extend quantum-dot Kitaev-chain
platforms beyond the essentially free-fermion regime.\\

The main challenge of the resonator-mediated implementation is the attainable
interaction strength, which is limited by the achievable dot--resonator
coupling and resonator frequency. In addition, implementing several resonator
modes within the same quantum-dot array introduces practical challenges
associated with spatial crowding, capacitive cross-talk, and hybridization
between nearby modes. These effects require careful engineering of the
resonator frequencies and couplings. One possible strategy is to exploit
different capacitive couplings and gate lever arms, allowing distinct
resonator modes to couple differently to the dots while maintaining the
desired effective interaction strengths. More generally, disorder in the dot
parameters, finite temperature and calibration of the critical manifold in quantum-dot arrays will place
additional constraints on the precision with which continuously varying
critical exponents can be extracted experimentally. Realistic electron temperatures, typically  in the range \cite{Dvir_2023,Bordin_2025}
$T_{\rm e}\sim 30$--$50\,\mathrm{mK}$, are not expected to pose a
limitation for the finite systems considered here, with
$L=8,10,12,14$. For these system sizes, the finite-size excitation gap
remains substantially larger than the thermal energy scale $k_{\rm B}T$,
such that thermal occupation of excited states is strongly suppressed.\\

Despite these challenges, the required observables are accessible already in
finite systems, making quantum-dot arrays a promising platform for studying
continuously varying quantum criticality without requiring the thermodynamic
limit. Beyond the specific Ashkin--Teller and eight-vertex realizations
considered here, the underlying strategy of combining gate-tunable
quantum-dot arrays with engineered resonator-mediated or electrostatic
interactions provides a flexible route toward a broader class of interacting
one-dimensional quantum systems. In this sense, the proposed architectures
offer not only a route to experimentally test continuously varying critical
behavior, but also a general framework for extending quantum-dot platforms toward strongly interacting quantum
many-body physics.
           
\section{Acknowledgements}
WM thanks Anasua Chatterjee and Miguel C. Belo for useful discussions on the experimental feasibility of the proposed resonator-based setup and Nick van Loo on measuring local parities. This work was supported by the Netherlands Organization for Scientific Research (NWO/OCW), as part of Quantum Limits (project number SUMMIT.1.1016). Computational resources were provided by the DelftBlue and Snellius high-performance computing facilities. NC acknowledges the financial support from the Royal Society (grant number URFR1251326). The code used to generate the numerical results presented in this work is available on Zenodo~\cite{zenodo_22816410}.
\newpage

\bibliography{apssamp}

@article{Kitaev_2001,
   title={Unpaired Majorana fermions in quantum wires},
   volume={44},
   ISSN={1468-4780},
   url={http://dx.doi.org/10.1070/1063-7869/44/10S/S29},
   DOI={10.1070/1063-7869/44/10s/s29},
   number={10S},
   journal={Physics-Uspekhi},
   publisher={Uspekhi Fizicheskikh Nauk (UFN) Journal},
   author={Kitaev, A Yu},
   year={2001},
   month=Oct, pages={131–136} }

@article{PhysRevB.110.125113,
  title = {Resolving chiral transitions in one-dimensional Rydberg arrays with quantum Kibble-Zurek mechanism and finite-time scaling},
  author = {Soto Garcia, Jose and Chepiga, Natalia},
  journal = {Phys. Rev. B},
  volume = {110},
  issue = {12},
  pages = {125113},
  numpages = {11},
  year = {2024},
  month = {Sep},
  publisher = {American Physical Society},
  doi = {10.1103/PhysRevB.110.125113},
  url = {https://link.aps.org/doi/10.1103/PhysRevB.110.125113}
}

@ARTICLE{2019Natur.568..207K,
       author = {{Keesling}, Alexander and {Omran}, Ahmed and {Levine}, Harry and {Bernien}, Hannes and {Pichler}, Hannes and {Choi}, Soonwon and {Samajdar}, Rhine and {Schwartz}, Sylvain and {Silvi}, Pietro and {Sachdev}, Subir and {Zoller}, Peter and {Endres}, Manuel and {Greiner}, Markus and {Vuleti{\'c}} and {}, Vladan and {Lukin}, Mikhail D.},
        title = "{Quantum Kibble-Zurek mechanism and critical dynamics on a programmable Rydberg simulator}",
      journal = {\nat},
         year = 2019,
        month = apr,
       volume = {568},
       number = {7751},
        pages = {207-211},
          doi = {10.1038/s41586-019-1070-1},
archivePrefix = {arXiv},
       eprint = {1809.05540},
 primaryClass = {quant-ph},
       adsurl = {https://ui.adsabs.harvard.edu/abs/2019Natur.568..207K}
}

@article{Leijnse_2012,
   title={Parity qubits and poor man’s Majorana bound states in double quantum dots},
   volume={86},
   ISSN={1550-235X},
   url={http://dx.doi.org/10.1103/PhysRevB.86.134528},
   DOI={10.1103/physrevb.86.134528},
   number={13},
   journal={Physical Review B},
   publisher={American Physical Society (APS)},
   author={Leijnse, Martin and Flensberg, Karsten},
   year={2012},
   month=Oct }

@article{Liu_2025,
   title={Scaling up a sign-ordered Kitaev chain without magnetic flux control},
   volume={7},
   ISSN={2643-1564},
   url={http://dx.doi.org/10.1103/PhysRevResearch.7.L012045},
   DOI={10.1103/physrevresearch.7.l012045},
   number={1},
   journal={Physical Review Research},
   publisher={American Physical Society (APS)},
   author={Liu, Chun-Xiao and Miles, Sebastian and Bordin, Alberto and ten Haaf, Sebastiaan L. D. and Mazur, Grzegorz P. and Bozkurt, A. Mert and Wimmer, Michael},
   year={2025},
   month=Feb }

@article{Bordin_2025,
   title={Enhanced Majorana stability in a three-site Kitaev chain},
   volume={20},
   ISSN={1748-3395},
   url={http://dx.doi.org/10.1038/s41565-025-01894-4},
   DOI={10.1038/s41565-025-01894-4},
   number={6},
   journal={Nature Nanotechnology},
   publisher={Springer Science and Business Media LLC},
   author={Bordin, Alberto and Liu, Chun-Xiao and Dvir, Tom and Zatelli, Francesco and ten Haaf, Sebastiaan L. D. and van Driel, David and Wang, Guanzhong and van Loo, Nick and Zhang, Yining and Wolff, Jan Cornelis and Van Caekenberghe, Thomas and Badawy, Ghada and Gazibegovic, Sasa and Bakkers, Erik P. A. M. and Wimmer, Michael and Kouwenhoven, Leo P. and Mazur, Grzegorz P.},
   year={2025},
   month=Mar, pages={726–731} }

@misc{aoun2023phasediagramashkintellermodel,
      title={Phase diagram of the Ashkin-Teller model}, 
      author={Yacine Aoun and Moritz Dober and Alexander Glazman},
      year={2023},
      eprint={2301.10609},
      archivePrefix={arXiv},
      primaryClass={math.PR},
      url={https://arxiv.org/abs/2301.10609}, 
}

@article{Liu_2022,
  title = {Tunable Superconducting Coupling of Quantum Dots via Andreev Bound States in Semiconductor-Superconductor Nanowires},
  author = {Liu, Chun-Xiao and Wang, Guanzhong and Dvir, Tom and Wimmer, Michael},
  journal = {Phys. Rev. Lett.},
  volume = {129},
  issue = {26},
  pages = {267701},
  numpages = {8},
  year = {2022},
  month = {Dec},
  publisher = {American Physical Society},
  doi = {10.1103/PhysRevLett.129.267701},
  url = {https://link.aps.org/doi/10.1103/PhysRevLett.129.267701}
}

@article{PhysRevLett541539,
  title = {Phase Diagram of Selenium Adsorbed on the Ni(100) Surface: A Physical Realization of the Ashkin-Teller Model},
  author = {Bak, Per and Kleban, P. and Unertl, W. N. and Ochab, J. and Akinci, G. and Bartelt, N. C. and Einstein, T. L.},
  journal = {Phys. Rev. Lett.},
  volume = {54},
  issue = {14},
  pages = {1539--1542},
  numpages = {0},
  year = {1985},
  month = {Apr},
  publisher = {American Physical Society},
  doi = {10.1103/PhysRevLett.54.1539},
  url = {https://link.aps.org/doi/10.1103/PhysRevLett.54.1539}
}

@article{Chepiga_2021,
   title={Kibble-Zurek exponent and chiral transition of the period-4 phase of Rydberg chains},
   volume={12},
   ISSN={2041-1723},
   url={http://dx.doi.org/10.1038/s41467-020-20641-y},
   DOI={10.1038/s41467-020-20641-y},
   number={1},
   journal={Nature Communications},
   publisher={Springer Science and Business Media LLC},
   author={Chepiga, Natalia and Mila, Frédéric},
   year={2021},
   month=Jan }

@article{PhysRevLett26832,
  title = {Eight-Vertex Model in Lattice Statistics},
  author = {Baxter, R. J.},
  journal = {Phys. Rev. Lett.},
  volume = {26},
  issue = {14},
  pages = {832--833},
  numpages = {0},
  year = {1971},
  month = {Apr},
  publisher = {American Physical Society},
  doi = {10.1103/PhysRevLett.26.832},
  url = {https://link.aps.org/doi/10.1103/PhysRevLett.26.832}
}

@article{Ashkin1943,
  title   = {Statistics of Two-Dimensional Lattices with Four Components},
  author  = {Ashkin, J. and Teller, E.},
  journal = {Physical Review},
  volume  = {64},
  pages   = {178},
  year    = {1943},
  doi     = {10.1103/PhysRev.64.178}
}

@book{Baxter1982,
  author    = {Baxter, Rodney J.},
  title     = {Exactly Solved Models in Statistical Mechanics},
  publisher = {Academic Press},
  address   = {London},
  year      = {1982},
  isbn      = {9780486462714},
  doi       = {10.1142/9789814415255_0002}
}

@article{Baxter1972wg,
    author = "Baxter, Rodney J.",
    title = "{Eight vertex model in lattice statistics and one-dimensional anisotropic Heisenberg chain. 1. Some fundamental eigenvectors}",
    doi = "10.1016/0003-4916(73)90439-9",
    journal = "Annals Phys.",
    volume = "76",
    pages = "1--24",
    year = "1973"
}

@article{Chepiga_2023,
   title={Eight-vertex criticality in the interacting Kitaev chain},
   volume={107},
   ISSN={2469-9969},
   url={http://dx.doi.org/10.1103/PhysRevB.107.L081106},
   DOI={10.1103/physrevb.107.l081106},
   number={8},
   journal={Physical Review B},
   publisher={American Physical Society (APS)},
   author={Chepiga, Natalia and Mila, Frédéric},
   year={2023},
   month=Feb }

@article{Zhang2025Floating,
  author  = {Zhang, Jin and Cant{\'u}, Sergio H. and Liu, Fangli
             and Bylinskii, Alexei and Braverman, Boris and Huber, Florian
             and Amato-Grill, Jesse and Lukin, Alexander and Gemelke, Nathan
             and Keesling, Alexander and Wang, Sheng-Tao
             and Meurice, Yannick and Tsai, Shan-Wen},
  title   = {Probing quantum floating phases in Rydberg atom arrays},
  journal = {Nature Communications},
  volume  = {16},
  pages   = {712},
  year    = {2025},
  doi     = {10.1038/s41467-025-55947-2}
}

@article{Hauke2016,
   title={Measuring multipartite entanglement through dynamic susceptibilities},
   volume={12},
   ISSN={1745-2481},
   url={http://dx.doi.org/10.1038/nphys3700},
   DOI={10.1038/nphys3700},
   number={8},
   journal={Nature Physics},
   publisher={Springer Science and Business Media LLC},
   author={Hauke, Philipp and Heyl, Markus and Tagliacozzo, Luca and Zoller, Peter},
   year={2016},
   month=Mar, pages={778–782} }

@article{PhysRevResearch7013215,
  title = {Numerical investigation of quantum phases and phase transitions in a two-leg ladder of Rydberg atoms},
  author = {Soto-Garcia, Jose and Chepiga, Natalia},
  journal = {Phys. Rev. Res.},
  volume = {7},
  issue = {1},
  pages = {013215},
  numpages = {15},
  year = {2025},
  month = {Feb},
  publisher = {American Physical Society},
  doi = {10.1103/PhysRevResearch.7.013215},
  url = {https://link.aps.org/doi/10.1103/PhysRevResearch.7.013215}
}

@article{PhysRevLett132076505,
  title = {Tunable Quantum Criticality in Multicomponent Rydberg Arrays},
  author = {Chepiga, Natalia},
  journal = {Phys. Rev. Lett.},
  volume = {132},
  issue = {7},
  pages = {076505},
  numpages = {5},
  year = {2024},
  month = {Feb},
  publisher = {American Physical Society},
  doi = {10.1103/PhysRevLett.132.076505},
  url = {https://link.aps.org/doi/10.1103/PhysRevLett.132.076505}
}

@misc{bordin2025probingmajoranalocalizationphasecontrolled,
      title={Probing Majorana localization of a phase-controlled three-site Kitaev chain with an additional quantum dot}, 
      author={Alberto Bordin and Florian J. Bennebroek Evertsz' and Bart Roovers and Juan D. Torres Luna and Wietze D. Huisman and Francesco Zatelli and Grzegorz P. Mazur and Sebastiaan L. D. ten Haaf and Ghada Badawy and Erik P. A. M. Bakkers and Chun-Xiao Liu and Ruben Seoane Souto and Nick van Loo and Leo P. Kouwenhoven},
      year={2025},
      eprint={2504.13702},
      archivePrefix={arXiv},
      primaryClass={cond-mat.mes-hall},
      url={https://arxiv.org/abs/2504.13702}, 
}

@article{PhysRevApplied12064049,
  title = {Measurements of Capacitive Coupling Within a Quadruple-Quantum-Dot Array},
  author = {Neyens, Samuel F. and MacQuarrie, E.R. and Dodson, J.P. and Corrigan, J. and Holman, Nathan and Thorgrimsson, Brandur and Palma, M. and McJunkin, Thomas and Edge, L.F. and Friesen, Mark and Coppersmith, S.N. and Eriksson, M.A.},
  journal = {Phys. Rev. Appl.},
  volume = {12},
  issue = {6},
  pages = {064049},
  numpages = {8},
  year = {2019},
  month = {Dec},
  publisher = {American Physical Society},
  doi = {10.1103/PhysRevApplied.12.064049},
  url = {https://link.aps.org/doi/10.1103/PhysRevApplied.12.064049}
}

@article{Dvir_2023,
   title={Realization of a minimal Kitaev chain in coupled quantum dots},
   volume={614},
   ISSN={1476-4687},
   url={http://dx.doi.org/10.1038/s41586-022-05585-1},
   DOI={10.1038/s41586-022-05585-1},
   number={7948},
   journal={Nature},
   publisher={Springer Science and Business Media LLC},
   author={Dvir, Tom and Wang, Guanzhong and van Loo, Nick and Liu, Chun-Xiao and Mazur, Grzegorz P. and Bordin, Alberto and ten Haaf, Sebastiaan L. D. and Wang, Ji-Yin and van Driel, David and Zatelli, Francesco and Li, Xiang and Malinowski, Filip K. and Gazibegovic, Sasa and Badawy, Ghada and Bakkers, Erik P. A. M. and Wimmer, Michael and Kouwenhoven, Leo P.},
   year={2023},
   month=Feb, pages={445–450} }

@article{Sau_2012,
   title={Realizing a robust practical Majorana chain in a quantum-dot-superconductor linear array},
   volume={3},
   ISSN={2041-1723},
   url={http://dx.doi.org/10.1038/ncomms1966},
   DOI={10.1038/ncomms1966},
   number={1},
   journal={Nature Communications},
   publisher={Springer Science and Business Media LLC},
   author={Sau, Jay D. and Sarma, S. Das},
   year={2012},
   month=jul }

@article{ten_Haaf_2024,
   title={A two-site Kitaev chain in a two-dimensional electron gas},
   volume={630},
   ISSN={1476-4687},
   url={http://dx.doi.org/10.1038/s41586-024-07434-9},
   DOI={10.1038/s41586-024-07434-9},
   number={8016},
   journal={Nature},
   publisher={Springer Science and Business Media LLC},
   author={ten Haaf, Sebastiaan L. D. and Wang, Qingzhen and Bozkurt, A. Mert and Liu, Chun-Xiao and Kulesh, Ivan and Kim, Philip and Xiao, Di and Thomas, Candice and Manfra, Michael J. and Dvir, Tom and Wimmer, Michael and Goswami, Srijit},
   year={2024},
   month=jun, pages={329–334} }

@article{PhysRevLett26834,
  title = {One-Dimensional Anisotropic Heisenberg Chain},
  author = {Baxter, R. J.},
  journal = {Phys. Rev. Lett.},
  volume = {26},
  issue = {14},
  pages = {834},
  numpages = {0},
  year = {1971},
  month = {Apr},
  publisher = {American Physical Society},
  doi = {10.1103/PhysRevLett.26.834},
  url = {https://link.aps.org/doi/10.1103/PhysRevLett.26.834}
}

@book{DiFrancesco1997,
    author = "Di Francesco, P. and Mathieu, P. and Senechal, D.",
    title = "{Conformal Field Theory}",
    doi = "10.1007/978-1-4612-2256-9",
    isbn = "978-0-387-94785-3, 978-1-4612-7475-9",
    publisher = "Springer-Verlag",
    address = "New York",
    series = "Graduate Texts in Contemporary Physics",
    year = "1997"
}

@article{PhysRevB245229,
  title = {Hamiltonian studies of the $d=2$ Ashkin-Teller model},
  author = {Kohmoto, Mahito and den Nijs, Marcel and Kadanoff, Leo P.},
  journal = {Phys. Rev. B},
  volume = {24},
  issue = {9},
  pages = {5229--5241},
  numpages = {0},
  year = {1981},
  month = {Nov},
  publisher = {American Physical Society},
  doi = {10.1103/PhysRevB.24.5229},
  url = {https://link.aps.org/doi/10.1103/PhysRevB.24.5229}
}

@article{PhysRevB76224423,
  title = {Scaling properties at the interface between different critical subsystems: The Ashkin-Teller model},
  author = {Lajk\'o, P\'eter and Turban, Lo\"{\i}c and Igl\'oi, Ferenc},
  journal = {Phys. Rev. B},
  volume = {76},
  issue = {22},
  pages = {224423},
  numpages = {9},
  year = {2007},
  month = {Dec},
  publisher = {American Physical Society},
  doi = {10.1103/PhysRevB.76.224423},
  url = {https://link.aps.org/doi/10.1103/PhysRevB.76.224423}
}

@book{giamarchi2004quantum,
  title     = {Quantum Physics in One Dimension},
  author    = {Giamarchi, T.},
  year      = {2004},
  publisher = {Clarendon Press},
  series    = {International Series of Monographs on Physics},
  isbn      = {9780198525004}
}

@article{Bravyi2011,
  author  = {Bravyi, Sergey and DiVincenzo, David P. and Loss, Daniel},
  title   = {Schrieffer--Wolff transformation for quantum many-body systems},
  journal = {Annals of Physics},
  volume  = {326},
  number  = {10},
  pages   = {2793--2826},
  year    = {2011},
  doi     = {10.1016/j.aop.2011.06.004}
}

@article{White1992,
  author  = {White, Steven R.},
  title   = {Density Matrix Formulation for Quantum Renormalization Groups},
  journal = {Physical Review Letters},
  volume  = {69},
  number  = {19},
  pages   = {2863--2866},
  year    = {1992},
  doi     = {10.1103/PhysRevLett.69.2863}
}

@article{OstlundRommer1995,
  author  = {{\"O}stlund, Stellan and Rommer, Stefan},
  title   = {Thermodynamic Limit of Density Matrix Renormalization},
  journal = {Physical Review Letters},
  volume  = {75},
  number  = {19},
  pages   = {3537--3540},
  year    = {1995},
  doi     = {10.1103/PhysRevLett.75.3537}
}

@article{Schollwock2011,
  author  = {Schollw{\"o}ck, Ulrich},
  title   = {The density-matrix renormalization group in the age of matrix product states},
  journal = {Annals of Physics},
  volume  = {326},
  number  = {1},
  pages   = {96--192},
  year    = {2011},
  doi     = {10.1016/j.aop.2010.09.012}
}

@article{Harvey_2018,
   title={Coupling two spin qubits with a high-impedance resonator},
   volume={97},
   ISSN={2469-9969},
   url={http://dx.doi.org/10.1103/PhysRevB.97.235409},
   DOI={10.1103/physrevb.97.235409},
   number={23},
   journal={Physical Review B},
   publisher={American Physical Society (APS)},
   author={Harvey, S. P. and Bøttcher, C. G. L. and Orona, L. A. and Bartlett, S. D. and Doherty, A. C. and Yacoby, A.},
   year={2018},
   month=jun }

@article{Fabrizio_2000,
   title={Critical properties of the double-frequency sine-Gordon model with applications},
   volume={580},
   ISSN={0550-3213},
   url={http://dx.doi.org/10.1016/S0550-3213(00)00247-9},
   DOI={10.1016/s0550-3213(00)00247-9},
   number={3},
   journal={Nuclear Physics B},
   publisher={Elsevier BV},
   author={Fabrizio, M. and Gogolin, A.O. and Nersesyan, A.A.},
   year={2000},
   month=Aug, pages={647–687} }

@article{Fulga_2013,
   title={Adaptive tuning of Majorana fermions in a quantum dot chain},
   volume={15},
   ISSN={1367-2630},
   url={http://dx.doi.org/10.1088/1367-2630/15/4/045020},
   DOI={10.1088/1367-2630/15/4/045020},
   number={4},
   journal={New Journal of Physics},
   publisher={IOP Publishing},
   author={Fulga, Ion C and Haim, Arbel and Akhmerov, Anton R and Oreg, Yuval},
   year={2013},
   month=Apr, pages={045020} }

@article{Senthil2004Deconfined,
  author  = {Senthil, T. and Vishwanath, Ashvin and Balents, Leon and Sachdev, Subir and Fisher, Matthew P. A.},
  title   = {Deconfined Quantum Critical Points},
  journal = {Science},
  volume  = {303},
  number  = {5663},
  pages   = {1490--1494},
  year    = {2004},
  doi     = {10.1126/science.1091806}
}

@article{Pronk2025Deconfined,
  author  = {Pronk, Niels T. and La Rivi{\`e}re, Bowy M. and Chepiga, Natalia},
  title   = {Deconfined quantum criticality in a frustrated Haldane chain with single-ion anisotropy},
  journal = {Physical Review B},
  volume  = {111},
  pages   = {L220412},
  year    = {2025},
  doi     = {10.1103/s53y-qmr4}
}

@article{PhysRevB521138,
  title = {Line of continuously varying criticality in the Ashkin-Teller quantum chain},
  author = {Yamanaka, Masanori and Kohmoto, Mahito},
  journal = {Phys. Rev. B},
  volume = {52},
  issue = {2},
  pages = {1138--1143},
  numpages = {0},
  year = {1995},
  month = {Jul},
  publisher = {American Physical Society},
  doi = {10.1103/PhysRevB.52.1138},
  url = {https://link.aps.org/doi/10.1103/PhysRevB.52.1138}
}

@article{PhysRevA81032334,
  title = {Entanglement and local extremes at an infinite-order quantum phase transition},
  author = {Rulli, C. C. and Sarandy, M. S.},
  journal = {Phys. Rev. A},
  volume = {81},
  issue = {3},
  pages = {032334},
  numpages = {7},
  year = {2010},
  month = {Mar},
  publisher = {American Physical Society},
  doi = {10.1103/PhysRevA.81.032334},
  url = {https://link.aps.org/doi/10.1103/PhysRevA.81.032334}
}

@article{FENDLEY1989549,
  author  = {Fendley, P. and Ginsparg, P.},
  title   = {Non-critical orbifolds},
  journal = {Nuclear Physics B},
  volume  = {324},
  number  = {3},
  pages   = {549--580},
  year    = {1989},
  doi     = {10.1016/0550-3213(89)90520-8}
}

@article{Delfino_2004,
   title={Universal ratios along a line of critical points. The Ashkin–Teller model},
   volume={682},
   ISSN={0550-3213},
   url={http://dx.doi.org/10.1016/j.nuclphysb.2004.01.007},
   DOI={10.1016/j.nuclphysb.2004.01.007},
   number={3},
   journal={Nuclear Physics B},
   publisher={Elsevier BV},
   author={Delfino, Gesualdo and Grinza, Paolo},
   year={2004},
   month=Mar, pages={521–550} }

@article{PhysRevB91165129,
  title = {Multiscale entanglement renormalization ansatz for spin chains with continuously varying criticality},
  author = {Bridgeman, Jacob C. and O'Brien, Aroon and Bartlett, Stephen D. and Doherty, Andrew C.},
  journal = {Phys. Rev. B},
  volume = {91},
  issue = {16},
  pages = {165129},
  numpages = {13},
  year = {2015},
  month = {Apr},
  publisher = {American Physical Society},
  doi = {10.1103/PhysRevB.91.165129},
  url = {https://link.aps.org/doi/10.1103/PhysRevB.91.165129}
}

@incollection{Cardy2006,
  author    = {Cardy, John},
  title     = {Boundary Conformal Field Theory},
  booktitle = {Encyclopedia of Mathematical Physics},
  editor    = {Fran{\c{c}}oise, Jean-Pierre and Naber, Gregory L. and Tsou, Sheung Tsun},
  publisher = {Academic Press},
  address   = {Oxford},
  pages     = {333--340},
  year      = {2006},
  doi       = {10.1016/B0-12-512666-2/00398-9}
}

@misc{zenodo_22816410,
  author    = {Missiaen, Warre},
  title     = {Code for {``Engineering weak universality with quantum dots''}},
  year      = {2026},
  publisher = {Zenodo},
  note      = {{DOI}: \mbox{10.5281/zenodo.22816410}}
}
\clearpage
\onecolumngrid
\appendix
\section{Mapping between the spin and Majorana representations of the Ashkin--Teller model}
In this section, we derive the equivalence between Eqs.~\eqref{eq:AT_spin} and \eqref{eq:AT_majorana} presented in the main text.

We consider two Majorana chains $\alpha=1,2$, with Majorana operators
$a_{\alpha,n}$ and $b_{\alpha,n}$ satisfying
\begin{equation*}
\{a_{\alpha,n},a_{\beta,m}\}
=
2\delta_{\alpha\beta}\delta_{nm},\quad \{b_{\alpha,n},b_{\beta,m}\}
=
2\delta_{\alpha\beta}\delta_{nm},\quad
\{a_{\alpha,n},b_{\beta,m}\}=0.
\end{equation*}

The Hamiltonian is
\begin{align*}
H_{\rm M}
={}&
t_1
\sum_{\alpha=1}^{2}\sum_n
i a_{\alpha,n}b_{\alpha,n}
+
t_2
\sum_{\alpha=1}^{2}\sum_n
i b_{\alpha,n}a_{\alpha,n+1}
\nonumber\\
&+
\lambda_x\sum_n
a_{1,n}b_{1,n}a_{2,n}b_{2,n}
\nonumber\\
&+
\lambda_z\sum_n
b_{1,n}a_{1,n+1}
b_{2,n}a_{2,n+1}.
\end{align*}

We now perform a Jordan--Wigner transformation separately along the two
chains. Since Pauli operators associated with different chains commute,
whereas Majorana operators belonging to different chains must
anticommute, we introduce Hermitian Klein factors $\eta_\alpha$
satisfying
\begin{equation*}
\eta_\alpha^\dagger=\eta_\alpha,
\qquad
\eta_\alpha^2=1,
\qquad
\{\eta_1,\eta_2\}=0,
\end{equation*}
and
\begin{equation*}
[\eta_\alpha,X_{\beta,n}]
=
[\eta_\alpha,Y_{\beta,n}]
=
[\eta_\alpha,Z_{\beta,n}]
=0.
\end{equation*}

For each chain, we define the Jordan--Wigner string
\begin{equation*}
\mathcal S_{\alpha,n}
=
\prod_{m<n}X_{\alpha,m},
\end{equation*}
and represent the Majorana operators as
\begin{equation*}
a_{\alpha,n}
=
\eta_\alpha\,
\mathcal S_{\alpha,n}
Z_{\alpha,n},
\qquad
b_{\alpha,n}
=
\eta_\alpha\,
\mathcal S_{\alpha,n}
Y_{\alpha,n}.
\end{equation*}

The Jordan--Wigner strings enforce the required anticommutation relations
between different sites of the same chain, while the Klein factors enforce
anticommutation between the two chains. For example, for
$\alpha\neq\beta$,
\begin{align*}
a_{\alpha,n}a_{\beta,m}
&=
\eta_\alpha\eta_\beta\,
\mathcal S_{\alpha,n}Z_{\alpha,n}
\mathcal S_{\beta,m}Z_{\beta,m}
\nonumber\\
&=
-\eta_\beta\eta_\alpha\,
\mathcal S_{\beta,m}Z_{\beta,m}
\mathcal S_{\alpha,n}Z_{\alpha,n}
\nonumber\\
&=
-a_{\beta,m}a_{\alpha,n},
\end{align*}
where we used the fact that spin operators belonging to different chains
commute and that
\begin{equation*}
\eta_\alpha\eta_\beta
=
-\eta_\beta\eta_\alpha.
\end{equation*}
The same argument applies to the remaining cross-chain anticommutation
relations.

We first consider the on-site Majorana bilinear. Using
\begin{equation*}
\mathcal S_{\alpha,n}^2=1,
\qquad
\eta_\alpha^2=1,
\end{equation*}
we obtain
\begin{align*}
i a_{\alpha,n}b_{\alpha,n}
&=
i\eta_\alpha^2
\mathcal S_{\alpha,n}^2
Z_{\alpha,n}Y_{\alpha,n}
\nonumber\\
&=
iZ_{\alpha,n}Y_{\alpha,n}.
\end{align*}
Since
\begin{equation*}
Z_{\alpha,n}Y_{\alpha,n}
=
-iX_{\alpha,n},
\end{equation*}
this gives
\begin{equation*}
i a_{\alpha,n}b_{\alpha,n}
=
X_{\alpha,n}.
\end{equation*}

Next, for neighboring sites,
\begin{equation*}
\mathcal S_{\alpha,n+1}
=
\mathcal S_{\alpha,n}X_{\alpha,n}.
\end{equation*}
Therefore,
\begin{align*}
i b_{\alpha,n}a_{\alpha,n+1}
&=
i\eta_\alpha^2
\mathcal S_{\alpha,n}Y_{\alpha,n}
\mathcal S_{\alpha,n+1}Z_{\alpha,n+1}
\nonumber\\
&=
iY_{\alpha,n}X_{\alpha,n}Z_{\alpha,n+1}.
\end{align*}
Using
\begin{equation*}
Y_{\alpha,n}X_{\alpha,n}
=
-iZ_{\alpha,n},
\end{equation*}
we find
\begin{equation*}
i b_{\alpha,n}a_{\alpha,n+1}
=
Z_{\alpha,n}Z_{\alpha,n+1}.
\end{equation*}

Thus, in the convention used here, the Jordan--Wigner transformation gives
\begin{equation*}
X_{\alpha,n}
=
i a_{\alpha,n}b_{\alpha,n},
\qquad
Z_{\alpha,n}Z_{\alpha,n+1}
=
i b_{\alpha,n}a_{\alpha,n+1}.
\end{equation*}

The quadratic part of the Hamiltonian consequently becomes
\begin{equation*}
H_0
=
t_1\sum_{\alpha,n}X_{\alpha,n}
+
t_2\sum_{\alpha,n}
Z_{\alpha,n}Z_{\alpha,n+1}.
\end{equation*}

We now consider the interaction proportional to $\lambda_x$. From
\begin{equation*}
i a_{\alpha,n}b_{\alpha,n}
=
X_{\alpha,n},
\end{equation*}
we have
\begin{equation*}
a_{\alpha,n}b_{\alpha,n}
=
-iX_{\alpha,n}.
\end{equation*}
Hence,
\begin{align*}
a_{1,n}b_{1,n}a_{2,n}b_{2,n}
&=
(-iX_{1,n})(-iX_{2,n})
\nonumber\\
&=
-X_{1,n}X_{2,n}.
\end{align*}
Therefore,
\begin{equation*}
\lambda_x\sum_n
a_{1,n}b_{1,n}a_{2,n}b_{2,n}
=
-\lambda_x\sum_n
X_{1,n}X_{2,n}.
\end{equation*}

Similarly,
\begin{equation*}
i b_{\alpha,n}a_{\alpha,n+1}
=
Z_{\alpha,n}Z_{\alpha,n+1}
\end{equation*}
implies
\begin{equation*}
b_{\alpha,n}a_{\alpha,n+1}
=
-iZ_{\alpha,n}Z_{\alpha,n+1}.
\end{equation*}
Consequently,
\begin{align*}
&b_{1,n}a_{1,n+1}
b_{2,n}a_{2,n+1}
\nonumber\\
&\qquad=
\left(-iZ_{1,n}Z_{1,n+1}\right)
\left(-iZ_{2,n}Z_{2,n+1}\right)
\nonumber\\
&\qquad=
-
Z_{1,n}Z_{1,n+1}
Z_{2,n}Z_{2,n+1}.
\end{align*}
Thus,
\begin{align*}
&\lambda_z\sum_n
b_{1,n}a_{1,n+1}
b_{2,n}a_{2,n+1}
\nonumber\\
&\qquad=
-\lambda_z\sum_n
Z_{1,n}Z_{1,n+1}
Z_{2,n}Z_{2,n+1}.
\end{align*}

Combining all contributions gives
\begin{align*}
H_{\rm M}
={}&
t_1\sum_{\alpha,n}X_{\alpha,n}
+
t_2\sum_{\alpha,n}
Z_{\alpha,n}Z_{\alpha,n+1}
\nonumber\\
&-
\lambda_x\sum_n
X_{1,n}X_{2,n}
\nonumber\\
&-
\lambda_z\sum_n
Z_{1,n}Z_{1,n+1}
Z_{2,n}Z_{2,n+1}.
\end{align*}

This is the quantum Ashkin--Teller Hamiltonian in the convention
\begin{equation*}
X_{\alpha,n}
=
i a_{\alpha,n}b_{\alpha,n},
\qquad
Z_{\alpha,n}Z_{\alpha,n+1}
=
i b_{\alpha,n}a_{\alpha,n+1}.
\end{equation*}

Note that the roles of \(X\) and \(Z\) may, in principle, be interchanged while leaving the quadratic sector equivalent to the Kitaev chain. In that convention, however, \(Z\) would correspond to the local fermion density of the Kitaev chain rather than to the magnetic order parameter of the Ashkin--Teller model.
The Klein factors are required at the level of the individual Majorana
operators to ensure the correct anticommutation relations between the two
chains. They do not appear in the final spin Hamiltonian because every
term contains an even number of Majorana operators from each chain, such
that the corresponding factors reduce to $\eta_\alpha^2=1$.
\section{Finite-size scaling details}
\label{app:fss}
In this section, we present the DMRG data used to generate Figs.~\ref{figATrange} and \ref{figXYZrange} of the main text.
\begin{table}[h]
\centering
\caption{Finite-size scaling fits of the Ashkin--Teller parity-string observables along the
self-dual critical line, $\lambda_x=\lambda_z=\lambda$, with
$t_1=t_2=-1$ and open boundary conditions.
Ground states were obtained using two-site DMRG with maximum bond dimension
$\chi_{\max}=800$, up to $80$ sweeps, truncation threshold
$\mathrm{svd}_{\min}=10^{-12}$, and energy-convergence criterion
$\Delta E<10^{-11}$.
The small-system fits use $L=8,10,12,14$, while the large-system fits use
$L=70,80,90,100$.
The scaling dimensions are extracted from
$|S_O(L)|=A_O D_c(L)^{-x_O}$, with
$D_c(L)=|(L+1)\sin[\pi L/2(L+1)]|$.
The quoted uncertainties are regression errors of the logarithmic fits.}
\label{tab:AT_BCFT_fits}
\begin{tabular}{|c|c|c|c|c|c|c|}
\hline
$\lambda$ & $x_Z^{\rm small}$ & $x_Z^{\rm large}$ & $x_Z^{\rm exact}$ & $x_P^{\rm small}$ & $x_P^{\rm large}$ & $x_P^{\rm exact}$ \\
\hline
0.00 & $0.12735\pm0.00004$ & $0.12569\pm0.00002$ & $0.12500$ & $0.25469\pm0.00008$ & $0.25138\pm0.00005$ & $0.25000$ \\
0.10 & $0.12550\pm0.00014$ & $0.12544\pm0.00001$ & $0.12500$ & $0.23756\pm0.00019$ & $0.23611\pm0.00004$ & $0.23501$ \\
0.20 & $0.12345\pm0.00023$ & $0.12512\pm0.00000$ & $0.12500$ & $0.22157\pm0.00030$ & $0.22232\pm0.00002$ & $0.22159$ \\
0.30 & $0.12124\pm0.00031$ & $0.12470\pm0.00001$ & $0.12500$ & $0.20641\pm0.00040$ & $0.20958\pm0.00000$ & $0.20938$ \\
0.40 & $0.11896\pm0.00037$ & $0.12411\pm0.00003$ & $0.12500$ & $0.19191\pm0.00049$ & $0.19752\pm0.00002$ & $0.19810$ \\
0.50 & $0.11677\pm0.00041$ & $0.12332\pm0.00005$ & $0.12500$ & $0.17801\pm0.00054$ & $0.18581\pm0.00005$ & $0.18750$ \\
0.60 & $0.11488\pm0.00042$ & $0.12232\pm0.00007$ & $0.12500$ & $0.16471\pm0.00056$ & $0.17416\pm0.00008$ & $0.17735$ \\
0.80 & $0.11332\pm0.00036$ & $0.12030\pm0.00008$ & $0.12500$ & $0.14001\pm0.00049$ & $0.15002\pm0.00011$ & $0.15720$ \\
\hline
\end{tabular}%
\end{table}
\begin{table}[h]
\centering
\caption{Finite-size scaling fits of the center Friedel-oscillation amplitude along the eight-vertex critical line $J_x=-J_z$ with fixed boundary conditions in the $Z$ direction ($h_B=5$). The scaling dimension $x_Z$ is extracted from $D(L)=A_DD_c(L)^{-x_Z}$, with
$D_c(L)=|(L+1)\sin[\pi L/2(L+1)]|$. The small-system fits use $L=8,10,12,14$ and were obtained using two-site DMRG with $\chi_{\max}=256$ and up to $60$ sweeps, while the large-system fits use $L=70,80,90,100$ with $\chi_{\max}=800$ and up to $80$ sweeps. In both cases, $\mathrm{svd}_{\min}=10^{-12}$ and the energy-convergence criterion is $\Delta E<10^{-11}$. The quoted uncertainties are regression errors of the logarithmic fits.}
\label{tab:XYZ_scaling}
\begin{tabular}{|c|c|c|c|}
\hline
$J_y/J_x$ & $x_Z^{\rm small}$ & $x_Z^{\rm large}$ & $x_Z^{\rm exact}$ \\
\hline
-0.40 & $0.19857 \pm 0.00112$ & $0.18660 \pm 0.00007$ & $0.18451$ \\
-0.20 & $0.23808 \pm 0.00267$ & $0.22041 \pm 0.00009$ & $0.21795$ \\
0.00 & $0.27624 \pm 0.00453$ & $0.25276 \pm 0.00010$ & $0.25000$ \\
0.20 & $0.31496 \pm 0.00637$ & $0.28500 \pm 0.00011$ & $0.28205$ \\
0.40 & $0.35649 \pm 0.00797$ & $0.31858 \pm 0.00013$ & $0.31549$ \\
0.60 & $0.40349 \pm 0.00919$ & $0.35588 \pm 0.00017$ & $0.35242$ \\
0.80 & $0.45901 \pm 0.00987$ & $0.40252 \pm 0.00024$ & $0.39758$ \\
\hline
\end{tabular}
\end{table}
\end{document}